\documentclass[preprint,12pt,authoryear]{elsarticle}
\usepackage{amssymb}
\usepackage[utf8]{inputenc}   
\graphicspath{ {./figures/} }
\usepackage{hyperref}
\usepackage{floatrow}
\usepackage{verbatim} %comments
\usepackage{amsfonts}
\usepackage{booktabs}
\usepackage{multirow}
\usepackage{threeparttable}
\usepackage{tabularx}
\usepackage{makecell}
\usepackage{hyperref}
\usepackage{lineno}
\usepackage{graphicx} 
\usepackage{subfig}      
\usepackage{algorithmic}
\usepackage{algorithm}
\usepackage{amsmath}
\usepackage{float}
\usepackage{setspace}
\floatstyle{plain}  
\restylefloat{figure}
\restylefloat{figure}
\restylefloat{table}

\journal{Knowledge-Based Systems}

\begin{document}

\floatsetup[table]{capposition=top}

\begin{frontmatter}

\title{Untrained Filtering with Trained Focusing for Superior Quantum Architecture Search}

\author[lable1]{Lian-Hui Yu}
\author[lable2]{Xiao-Yu Li$^\dagger$}
\author[lable3]{Geng Chen}
\author[lable1,lable5]{Qin-Sheng Zhu}
\author[lable4]{Hui Li}
\author[lable3]{Guo-Wu Yang}

\affiliation[lable1]{organization={School of Physics, University of Electronic Science and Technology of China},
            city={Cheng Du},
            postcode={610054}, 
            state={Si Chuan},
            country={China}}
\affiliation[lable2]{organization={School of Information and Software Engineering, University of Electronic Science and Technology of China},
	city={Cheng Du},
	postcode={610054}, 
	state={Si Chuan},
	country={China}}
\affiliation[lable3]{organization={School of Computer Science and Engineering, University of Electronic Science and Technology of China},
	city={Cheng Du},
	postcode={610054}, 
	state={Si Chuan},
	country={China}}
\affiliation[lable4]{organization={JILA and Department of Physics, University of  Colorado},
	city={Boulder},
	postcode={80309-0440}, 
	state={Colorado},
	country={USA}}
\affiliation[lable5]{organization={Advanced Cryptography and System Security Key Laboratory of Sichuan Province},
	city={Cheng Du},
	postcode={610103}, 
	state={Si Chuan},
	country={China}}

\begin{abstract}
Quantum architecture search (QAS) represents a fundamental challenge in quantum machine learning. Unlike previous methods that treat it as a static search process, from a perspective on QAS as an item retrieval task in vast search space, we decompose the search process into dynamic alternating phases of coarse and fine-grained knowledge learning. We propose $\textbf{Q}$uantum $\textbf{U}$ntrained-$\textbf{E}$xplored $\textbf{S}$ynergistic $\textbf{T}$rained $\textbf{A}$rchitecture (QUEST-A),a framework through coarse-grained untrained filtering for rapid search space reduction and fine-grained trained focusing for precise space refinement in progressive QAS. QUEST-A develops an evolutionary mechanism with knowledge accumulation and reuse to enhance multi-level knowledge transfer in architecture searching. Experiments demonstrate QUEST-A's superiority over existing methods: enhancing model expressivity in signal representation, maintaining high performance across varying complexities in image classification, and achieving order-of-magnitude precision improvements in variational quantum eigensolver tasks, providing a transferable methodology for QAS.
\end{abstract}


\end{frontmatter}

\section{Introduction}
\label{introduction}

Quantum Machine Learning (QML) aims to reinterpret and enhance classical machine learning within the framework of quantum computing \cite{biamonte2017quantum}. QML not only inherits the wide applicability of traditional machine learning, encompassing diverse tasks from classification \cite{huang2021power,wang2022quantumnas,anagolum2024elivagar} and regression to generative modeling \cite{niu2022entangling,zoufal2021variational} and reinforcement learning \cite{huang2022quantum}, but also incorporates the unique advantages of quantum computing. In the era of Noisy Intermediate-Scale Quantum (NISQ) devices \cite{preskill2018quantum}, QML has garnered significant attention for its potential to solve complex real-world problems.

Currently, Variational Quantum Circuits (VQCs) represent the most widely adopted method in QML \cite{gong2024quantum}. VQCs optimize model performance through iterative optimization of quantum circuit parameters until the quantum circuit can learn and represent the input problem with high accuracy \cite{thanasilp2023subtleties}. From this perspective, QML can be abstracted as a architecture optimization problem guided by a loss function and subject to multiple quantum constraints. This perspective, though radical, bridges the gap between classical machine learning and its quantum counterpart. While VQCs provide a powerful framework for quantum machine learning, designing efficient quantum circuits faces significant challenges as problem scales increase. Traditional manual design methods, although highly task-specific, often struggle to handle complex, large-scale problems effectively. Conversely, the adoption of generalized template-based quantum circuit designs, while simplifying the design process, frequently leads to severe performance bottlenecks, particularly when confronted with issues such as barren plateaus \cite{mcclean2018barren}.

To address these challenges, Quantum architecture search (QAS) has emerged as a promising solution \cite{wang2022quantumnas,du2022quantum,zhang2022differentiable,wu2023quantumdarts,he2024training}. Similar to Neural Architecture Search (NAS), QAS can be viewed as an item retrieval task within a vast architecture space, where the core challenge lies in efficiently retrieving target quantum circuit structures with limited prior knowledge of the search objective. This highlights the importance of dynamic knowledge accumulation during the search process.

Currently, most QAS methods draw inspiration from the NAS paradigm and can be broadly categorized into heuristic, differential-based, and sampling-based approaches based on their optimization strategies. Heuristic algorithms \cite{wang2022quantumnas,williams1998automated,lamata2018quantum,yu2024qusl} address the circuit architecture optimization problem through quadratic modeling and search the solution space according to predefined rules. These methods offer strong interpretability, providing clear insights into the optimization process. However, they suffer from limited scalability and can only probabilistically obtain global optima, often converging to local optima in large-scale problems. As an extension of variational quantum algorithms, differential-based QAS \cite{zhang2022differentiable,wu2023quantumdarts,zhang2021neural,du2022quantum,ma2024continuous} constructs end-to-end differentiable workflows, utilizing gradient-guided searches with predefined supercircuits or modules. This approach effectively leverages gradient information; nevertheless, its search efficiency is constrained by preset structures and optimization techniques, resulting in computationally intensive training processes. Training-free QAS \cite{anagolum2024elivagar,he2024training}, an emerging efficient search mode derived from classical training-free architecture search methods \cite{zhou2022training}, employs intrinsic characteristics of quantum circuits as performance proxy indicators to rapidly screen candidate circuit sets. The strength of this approach lies in its search efficiency, but rapid, undirected nature of the search also risks overlooking potentially superior architectural configurations. These methods essentially view QAS as a static search problem in a huge search space, using fixed retrieval strategies for either coarse-grained rapid pruning or fine-grained directed traversal, which explains their difficulty in balancing efficiency and accuracy.

\begin{figure}[!t]
	\centering
	\subfloat[]{
		\includegraphics[width=0.35\textwidth,trim=8cm 8cm 8cm 7.5cm,clip]{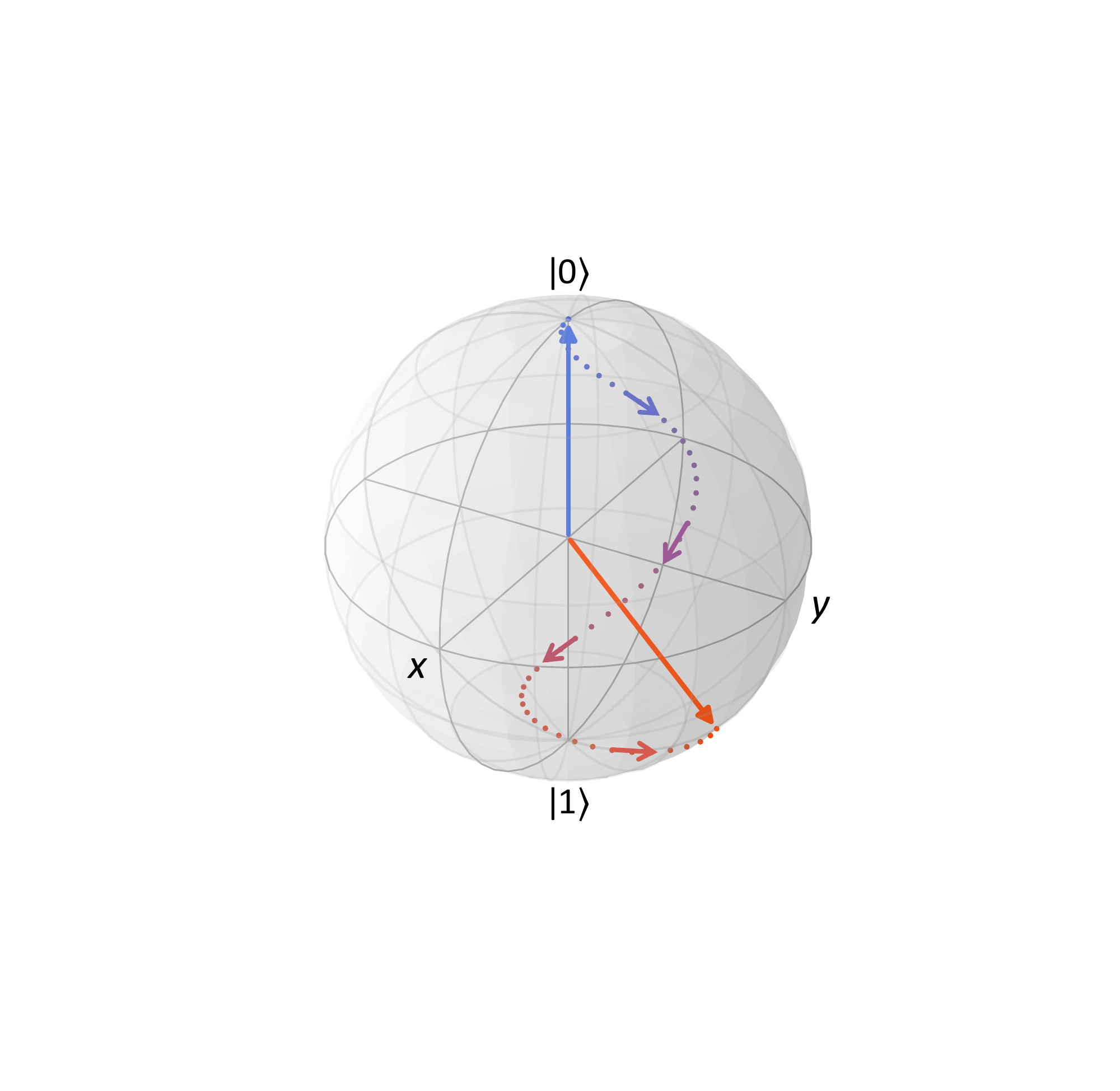}
		\label{fig:subfig_a}
	}
	\quad
	\subfloat[]{
		\includegraphics[width=0.35\textwidth,trim=8cm 8cm 8cm 7.5cm,clip]{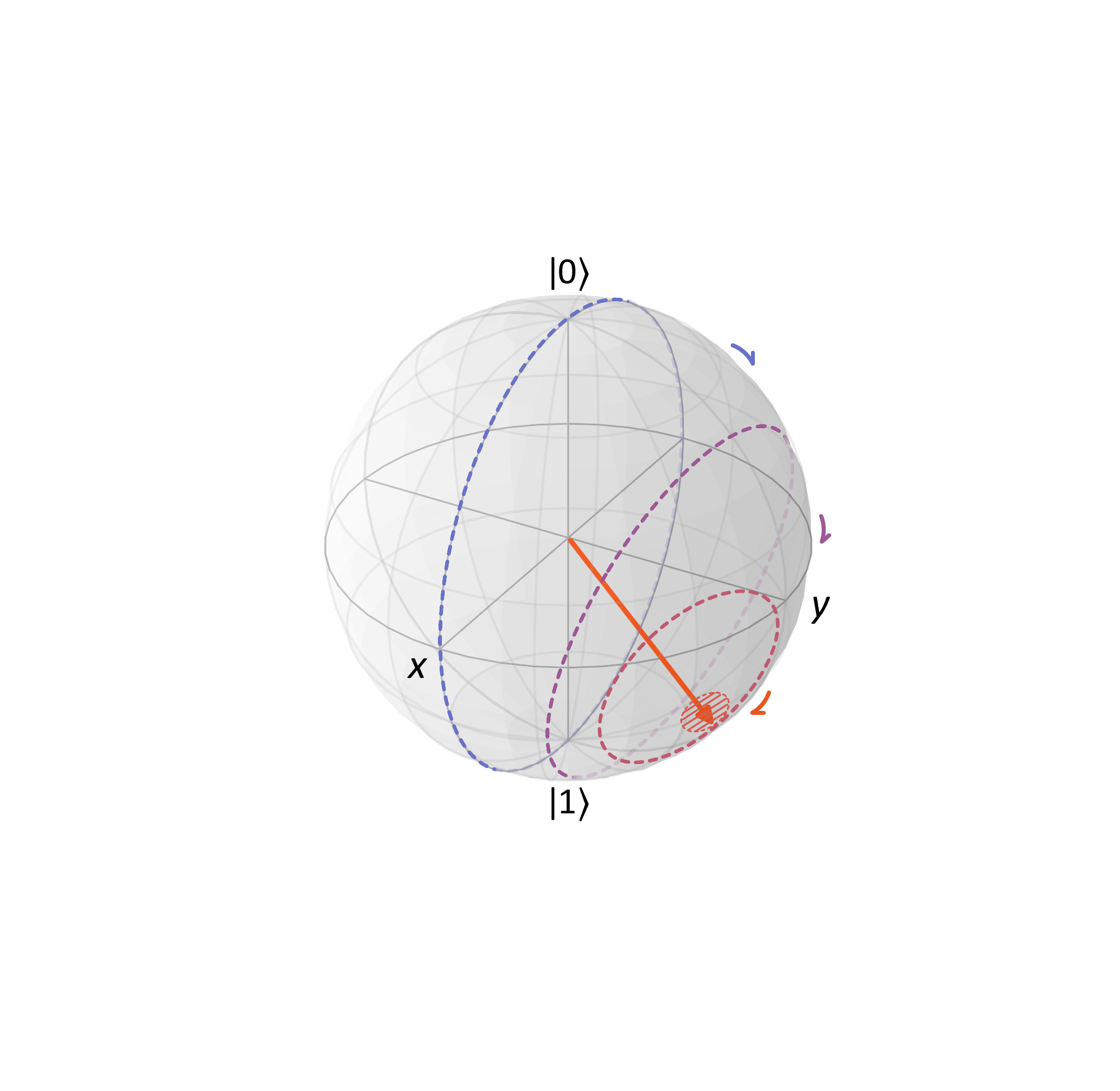}
		\label{fig:subfig_b}
	}
	\caption{
		\footnotesize
		Comparison between gradient-based and training-free QAS methods. Output single-qubit states are used to visualize the circuit search process. (a) Continuous gradient-based approach gradually converges to target circuit through gradient guidance. (b) Discrete training-free approach narrows search space through screening indicators to locate target circuit vicinity.
		}
	\label{fig:comparison}
\end{figure}

Considering the respective strengths and limitations of the aforementioned QAS methodologies, our work is motivated by the need to incorporate dynamic knowledge accumulation for progressive QAS. To this end, we propose the QUEST-A, which aims to surmount the limitations of singular paradigms by synergistically combining coarse-grained untrained filtering and fine-grained trained focusing. Specifically, QUEST-A iteratively combines coarse-grained untrained filtering for rapid search space reduction with fine-grained trained focusing for precise space refinement in a dynamic search process. Based on the flexibility of reupload circuits, the multi-level strategy naturally forms an iterative evolutionary process, where knowledge accumulation and reuse strategies unify coarse-grained rapid pruning and fine-grained precise search of the architecture space. This dynamic process emphasizes comprehensive knowledge utilization during search, avoiding the partial understanding of QAS in previous methods.

The contributions of our work are as follows: 

\begin{enumerate}
	\item Formulated QAS as an item retrieval task with dynamic knowledge learning, and developed a framework that iteratively combines coarse-grained untrained filtering for rapid search space reduction with fine-grained trained focusing for precise space refinement.
	\item Developed an evolutionary mechanism with knowledge accumulation and reuse strategies, integrating coarse-grained untrained filtering and fine-grained trained focusing through genetic operations and fitness-guided selection, enabling knowledge transfer and accumulation during searching.
	\item In numerical simulations and NISQ environment experiments across multiple types of tasks, QUEST-A demonstrated performance superior to state-of-the-art methods. Specifically, it achieved up to $81.8\%$ performance improvement in representation tasks, maintained over $8 5\%$ high accuracy in $10$-class image classification tasks, and enhanced precision by up to $11$ orders of magnitude in VQE tasks.
\end{enumerate}

\newpage

\section{Related Works}
\label{Related Works}

\subsection{Quantum Circuits for Quantum Machine Learning}
\label{Quantum circuits for quantum machine learning}

\begin{figure}[h]
	\hspace{0cm}
	\includegraphics[width=0.85\textwidth, trim=9cm 8cm 8cm 8.2cm,clip]{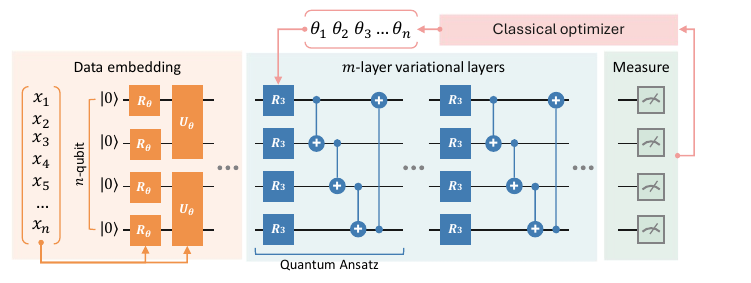}
	\caption{
		\footnotesize
		Quantum-classical hybrid machine learning framework. }
	\label{fig_VQC}
\end{figure}

VQCs have emerged as the dominant methodology in QML. VQCs comprise modular circuits, encompassing data encoding, trainable variational gates, and measurement operations, thereby forming a quantum-classical hybrid computational process as show in Fig.~\ref{fig_VQC}.

This process initiates with the encoding of classical data into quantum states, followed by the manipulation of these states through variational and entangling gates, and culminates in the conversion of results back to classical data via measurement. The training process of VQCs involves parameter optimization, typically employing gradient descent methods. From a knowledge learning perspective, gradient computation in quantum circuits provides fine-grained knowledge about circuit parameters, costs scaling linearly with the number of parameters. This characteristic, combined with the noise sensitivity of NISQ devices\cite{preskill2018quantum}, underscores the significant impact of circuit structure on QML performance.

The influence of quantum circuit structure on learning performance is primarily manifested in three aspects: circuit scale, training potential, and hardware adaptability. The circuit scale, one of the most concerned properties in the NISQ era, is determined by the type, quantity, and arrangement of gates, which can be evaluated through inherent characteristics \cite{sim2019expressibility}. Training potential reflects the trade-off between expressibility and the number of trainable parameters $-$ complex circuits do not necessarily enhance expressibility but often increase learning costs. In the NISQ era, circuit design can also consider hardware connectivity constraints and noise characteristics.

\subsection{Quantum Architecture Search}
\label{Quantum Architecture Search}

QAS is an item retrieval task aimed at finding optimal quantum circuit structures and parameters within a large-scale search space. The complexity of QAS can be quantified by analyzing its search space: as described in Fig.~\ref{fig_VQC}, consider a quantum circuit model comprising $n$ qubits and m layers, where each layer can select from a set of quantum gates $G$ with cardinality $k = |G|$. Under this model, the search space size reaches $O(k^{n×m})$ \cite{ge2024quantum}. This exponential growth indicates that the computational complexity of QAS increases significantly as the number of qubits and circuit depth increase. Furthermore, QAS involves both discrete structure search and continuous parameter optimization \cite{zhang2022differentiable}, further augmenting the problem's complexity.

Although the precise computational complexity classification of QAS remains an open question, there are substantial grounds to conjecture that its complexity is at least comparable to NP-hard problems \cite{wu2023quantumdarts,diez2021quantum}. The QAS problem can be formally described as follows: Let the candidate quantum gates be represented by matrix $M \in G^{n\times m}$, define $U_{ij} = \sigma(M_{ij})$, where $\sigma$ maps the quantum gate on the $i$-th qubit and $j$-th layer to a $2^n \times 2^n$ unitary matrix. The unitary matrix corresponding to the entire quantum circuit can be expressed as follow

\begin{equation}
	U = \prod_{j=1}^m \prod_{i=1}^n U_{ij} = \prod_{j=1}^m \prod_{i=1}^n \sigma(M_{ij}),
\end{equation}

based on this, the search objective of QAS can be formalized as $\min_{M \in G^{n\times m}} L(U, D),$ where $L$ denotes the loss function, and $D$ represents the dataset or target unitary matrix for the specific problem. The core of QAS lies in efficiently retrieving the quantum circuit structure $M$ and its parameters that minimize the corresponding loss function within the $O(k^{n×m})$-scale search space with limited prior knowledge. It is noteworthy that $G^{n\times m}$ represents the set of possible quantum gate combinations, the selection of which directly impacts the size and complexity of the search space.

Meticulously designed quantum circuit templates can reduce the value of $k$, thereby diminishing the scale of the search space and making $O(k^{n×m})$ space more tractable. Consequently, the selection of high-quality quantum circuit templates as starting points for the search process becomes paramount.

\subsection{Ansatz for QAS}
\label{Ansatz for QAS}

In quantum computing, an ansatz refers to a parameterized quantum circuit structure. In QAS, ansatzes manifest as quantum circuit templates, defining the fundamental structure and characteristics of the search space.

The cascading CNOT circuit is currently the most commonly used ansatz \cite{mcclean2018barren}. By constructing complex entangled states, it establishes global correlations between qubits and serves as a prevalent template due to its universality. However, this structure assigns equal weights to all interactions, potentially obscuring feature importance \cite{wang2022quantumnas}, and the complex entanglement patterns increase the difficulty of parameter optimization. The advancement in understanding quantum circuit characteristics \cite{canatar2022bandwidth,cerezo2021cost,mcclean2018barren} has given rise to various novel ansatzes. Hardware-efficient ansatzes \cite{anagolum2024elivagar} consider the specific structure and gate sets of quantum hardware, while tensor network-inspired ansatzes \cite{lazzarin2022multi} leverage theoretical insights from classical tensor networks.

\begin{figure}[h]
	\hspace{0cm}
	\includegraphics[width=0.85\textwidth, trim=9cm 8cm 8cm 8.2cm,clip]{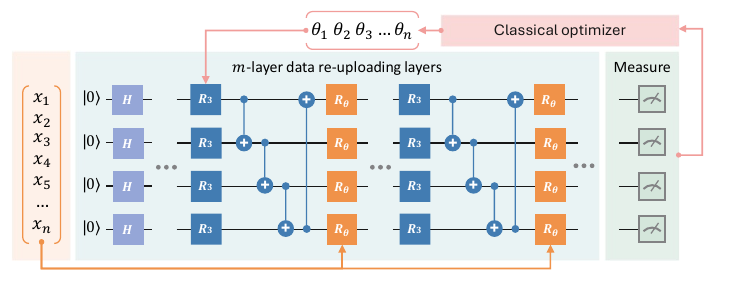}
	\caption{
		\footnotesize
		Hybrid machine learning framework based on re-uploading quantum circuits.}
	\label{DateReup}
\end{figure}

As an emerging quantum circuit ansatz, re-uploading circuits \cite{perez2020data} alternate data encoding and parameter variation processes, enabling multiple utilizations of input data and thus enhancing the circuit's expressivity. This design establishes a complex, data-oriented non-unitary process that can approximate any continuous function \cite{zhao2024quantum}, while providing increased flexibility and controllability in the search space. More importantly, by incorporating data encoding into the optimization scope \cite{ono2023demonstration}, re-uploading circuits' modular and repeatable structure provides an ideal template for efficient knowledge transfer and accumulation during architecture searching.

\section{Methodology}
\label{Methodology}

\subsection{Overview of QUEST-A}
\label{Overview of QUEST-A}

\begin{figure}[t]
	\hspace{0cm}
	\includegraphics[width=1.0\textwidth, trim=2.9cm 8.5cm 2.9cm 9cm,clip]{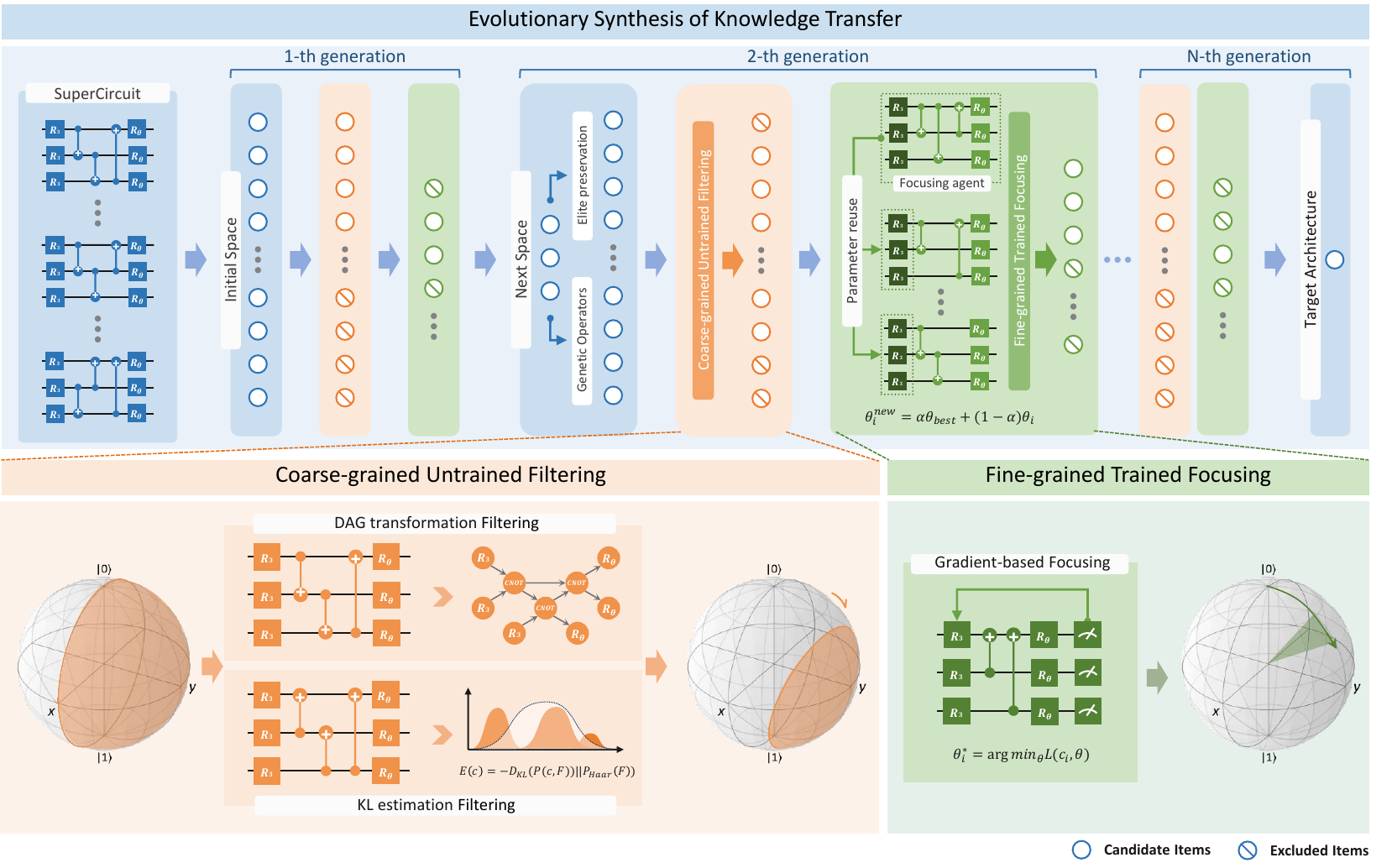}
	\caption{
		\footnotesize
		QUEST-A framework. }
	\label{framework}
\end{figure}

Traditional QAS approaches rely on fixed search strategies and face inherent difficulties in balancing search accuracy and efficiency. This limitation suggests the need for a dynamic knowledge-driven approach. A key insight of this study is the decomposition of QAS into two alternating phases: coarse-grained retrieval guided by circuit inherent properties, and fine-grained retrieval guided by task performance. This decomposition strategy provides a more tractable solution compared to directly addressing the entire search problem. As shown in Fig.~\ref{framework}, the QUEST-A framework consists of three core components corresponding to untrained filtering, trained focusing, and knowledge transfer strategies:

\begin{enumerate}
	\item $\textbf{Untrained Filtering:}$ This component constructs the initial search space using re$-$uploading circuit templates to indexes and manages candidate structures through supercircuit technology. The pruning process employs two complementary coarse$-$grained evaluation methods: first analyzing structural complexity using Directed Acyclic Graph (DAG) topological features, then calculating the expressive power of quantum state distributions via KL-divergence.
    \item $\textbf{Trained Focusing:}$ QUEST-A applies fine-grained search on filtered candidate spaces through gradient-guided mechanisms. This process progressively refines promising architectures in the reduced search space. The knowledge reuse strategy enables transfer of architectural knowledge between candidates, guiding the search towards superior regions.
    \item $\textbf{Evolutionary Synthesis of Knowledge Transfer:}$ QUEST-A employs evolutionary algorithms as a knowledge transfer framework, where each generation represents a specific region of the architecture space. Population initialization corresponds to candidate search regions generation in untrained filtering, fitness evaluation combines coarse-grained circuit properties and fine-grained task performance, selection operations are based on multi-level knowledge criteria, while crossover and mutation operations generate new search regions through structure manipulation.
\end{enumerate}

These three coupled components form a closed-loop search system. By alternating coarse-grained untrained filtering and fine-grained trained focusing within the evolutionary framework, QUEST-A achieves architecture discovery with progressive knowledge accumulation.

\subsection{Untrained Filtering}
\label{Untrained Filtering}

QAS can be considered a retrieval task, where appropriate reduction of the search space facilitates the knowledge learning process. Untrained filtering primarily focuses on discrete quantum circuit structures, thus further simplifying the problem into a discrete database retrieval task. Consequently, establishing an appropriate indexing mechanism is crucial to enable efficient knowledge discovery. Traditional supercircuit methods \cite{wang2022quantumnas,du2022quantum}, typically construct and store the entire supercircuit at once as indexing , resulting in substantial memory overhead and training costs.

In contrast, QUEST-A adopts a dynamic indexing strategy, effectively overcoming these limitations. Formally, given the number of qubits $N$, maximum circuit depth $L$, and a set of available quantum gates $G (|G| = Q)$, the indexer $A$ for the search space $S$ can be defined as $A: \{0, 1, ..., Q-1\}^{N×L} \rightarrow S$, where $\{0, 1, ..., Q-1\}^{N×L}$ represents the $N×L$-dimensional discrete index space. In the $t$-th iteration, the sampled specific quantum circuit structure $a^{t}$ can be expressed as $a^{t} = A(i^{t}), i^{t} \in \{0, 1, ..., Q-1\}^{N×L}$. This dynamic indexing method enhances memory efficiency by avoiding the need to store the entire supercircuit structure, while also improving computational efficiency by generating and evaluating specific structures only when necessary.

Furthermore, through the integration of re-uploading circuits in Section ~\ref{Ansatz for QAS}, QUEST-A enables the search for optimal data embedding methods \cite{perez2020data}. Specifically, this design partitions the overall encoding task into individual circuit modules, optimizing the embedding process for each module through QAS. 

Building upon the QUEST-A framework, the intrinsic properties of quantum circuits are further utilized to optimize the search process. These properties, directly related to the structural complexity and potential expressivity of the circuits, enable rapid identification and exclusion of candidate circuits with evident deficiencies in circuit scale and expressivity, thereby narrowing the search space \cite{he2024training}. QUEST-A employs two complementary evaluation methods: topological analysis based on Directed Acyclic Graphs (DAGs) and expressivity assessment using Kullback-Leibler (KL) divergence.

\begin{figure}[h]
	\hspace{0cm}
	\includegraphics[width=0.6\textwidth, trim=1cm 1.5cm 1cm 3cm,clip]{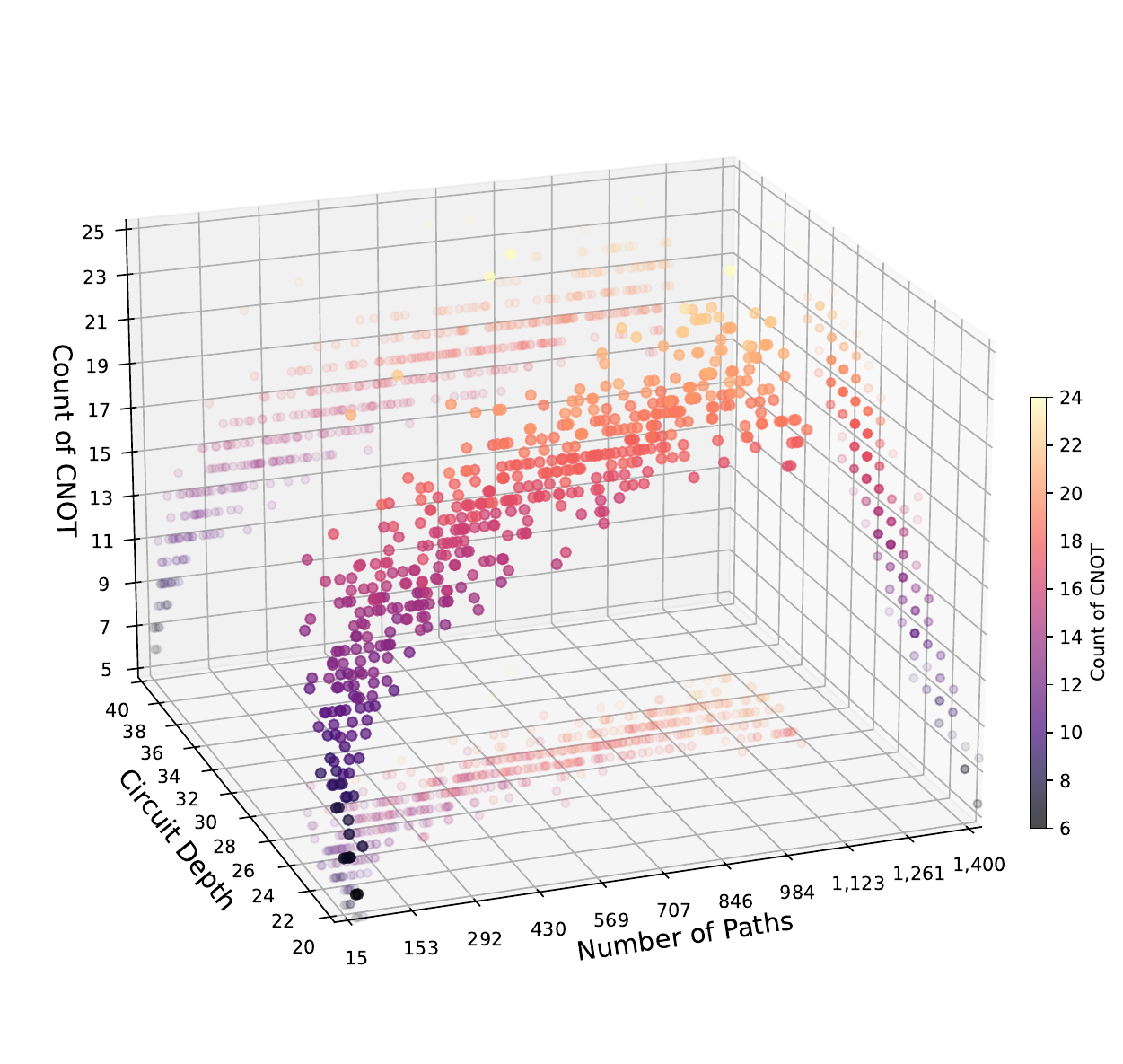}
	\caption{
		\footnotesize
		Relationship between DAG-path count and circuit complexity for $500$ random quantum circuits. As the DAG-path count increases, both circuit depth and number of CNOT show logarithmic growth, indicating a strong correlation between DAG-path count and circuit scale. }
	\label{DAGrelevance}
\end{figure}

Quantum circuits can be represented as DAGs \cite{holmes2022connecting}, where nodes represent quantum gate operations and edges represent qubit evolution paths. Based on this representation, the number of paths from input to output in the DAG is used as a proxy indicator for circuit complexity. For a given quantum circuit c, its path count is defined as $P(c)$. As demonstrated by the empirical study in Fig.~\ref{DAGrelevance}, $P(c)$ generally exhibits a positive correlation with the number of CNOT gates and circuit depth. This correlation highlights the effectiveness of $P(c)$ as an indicator of circuit complexity.

The use of $P(c)$ aims to balance between circuit expressivity and trainability, selecting circuits that potentially avoid both under-parameterization and the barren plateau problem associated with over-parameterization, as suggested by studies \cite{holmes2022connecting,cerezo2021cost} on quantum circuit complexity and trainability. This screening mechanism effectively eliminates candidate circuits that clearly do not meet task requirements. To complement the coarse-grained topological analysis, QUEST-A introduces an expressivity assessment $D(c)$ based on KL divergence. This method quantifies circuit expressivity by comparing the quantum state distribution generated by the circuit with the ideal Haar random distribution \cite{jiang2005maxima}. Specifically, for a quantum circuit c, its expressivity can be quantified by the following: 

\begin{equation}
E(c) = -D_{KL}(P(c, F) || P_{Haar}(F)),
\end{equation}

where $P(c, F)$ is the fidelity distribution of quantum states generated by circuit c, $P_{Haar}(F)$ is the Haar random distribution, and $D_{KL}$ denotes KL divergence. $F = |\langle\psi_\theta|\psi_\phi\rangle|^2$ defines the fidelity of quantum states, where $\theta$ and $\phi$ are the parameters of circuit c. While the exact value of quantum state fidelity can be calculated from the density matrices before and after quantum state evolution, in practice, $P(c, F)$ is approximated by random parameter sampling of circuit $c$. The analytical form of $P_{Haar}(F)$ is:

\begin{equation}
P_{Haar}(F) = (N - 1)(1 - F)^{N - 2},
\end{equation}

where $N$ is the dimension of the Hilbert space. A larger $E(c)$ value indicates higher expressivity, reflecting the circuit's ability to explore a larger portion of the Hilbert space.

Topological analysis, though effective for circuit complexity evaluation, cannot capture quantum-specific properties of circuits. KL divergence complements this limitation by providing a quantitative measure of circuit expressivity. This approach aligns with work research \cite{sim2019expressibility} on quantum circuit expressivity, emphasizing the importance of state distribution uniformity in evaluating circuit performance. In practical work, the relationship between KL divergence and candidate circuit performance does not exhibit a simple linear correlation. Instead, it provides a range of circuits with relatively stable performance, which can be used to filter out a series of circuit structures with lower training potential.

\begin{figure}[t]
	\hspace{0cm}
	\includegraphics[width=0.6\textwidth, trim=0cm 0cm 0cm 0cm,clip]{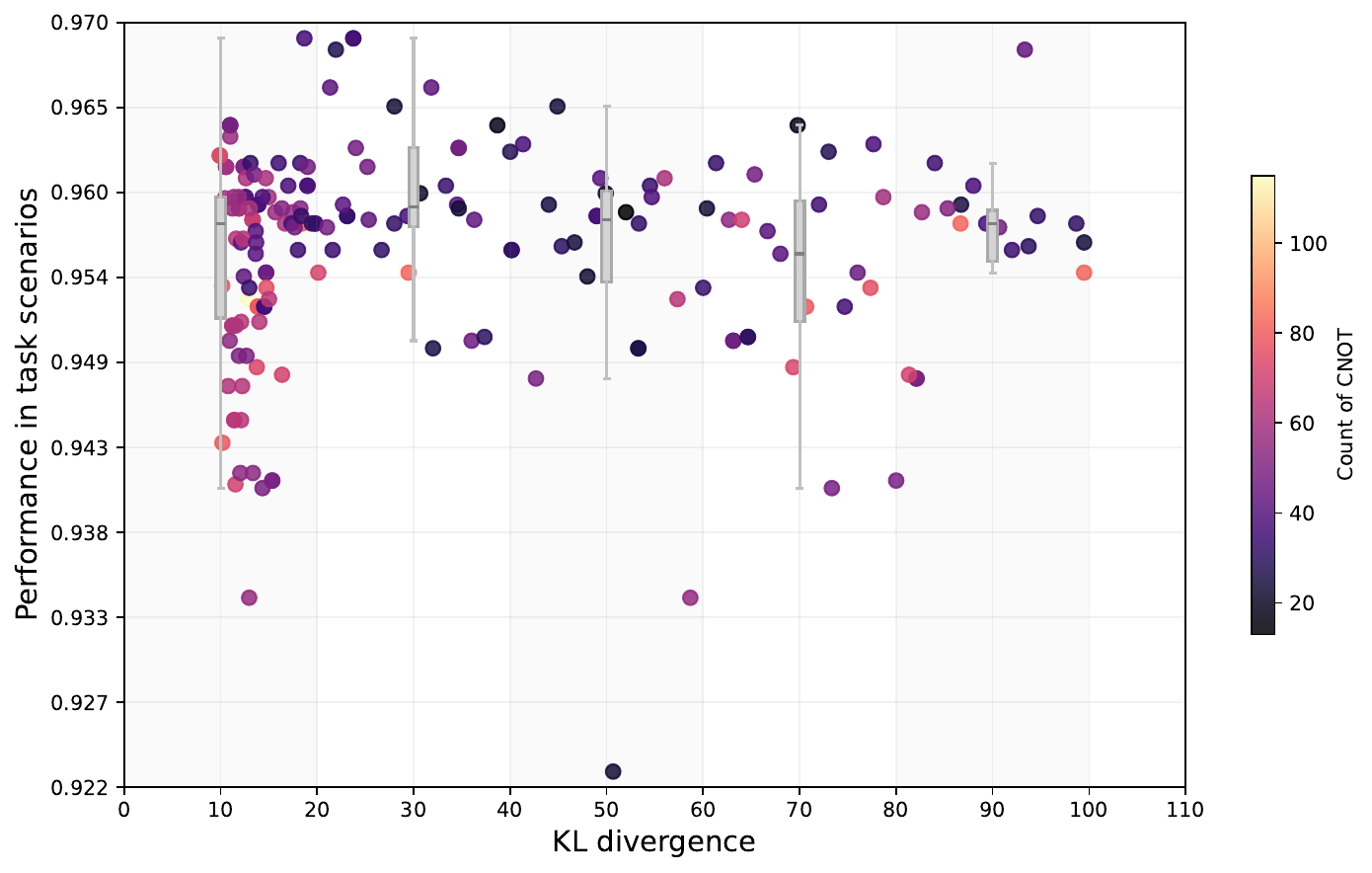}
	\caption{
		\footnotesize
		Relationship between KL divergence and quantum circuit performance in task scenarios.The test evaluates KL divergence for $200$ random circuits using data re-uploading circuits as templates. These circuits were trained on an MNIST $4$-class classification task until convergence. Outliers have been removed. }
	\label{KL}
\end{figure}

By combining $P(c)$ and $E(c)$, QUEST-A achieves rapid and relatively accurate screening of high-potential candidate circuits. The integration of these proxy indicators, based on topological complexity and expressive potential reflecting the multifaceted nature of quantum circuit performance, providing effective individuals for subsequent fine-tuning optimization.

\subsection{Trained Focusing}
\label{Trained Focusing}

After completing the untrained filtering, the QUEST-A framework enters the trained focusing phase. The primary objective of this stage is to perform fine-grained search in the reduced architecture space. Through computationally intensive methods, this phase aims to fully exploit the potential of each circuit structure.

In the field of QAS, gradient descent has been widely recognized as the most prevalent fine-grained optimization method. Given a set of candidate quantum circuits $\mathcal{P}'_g = \{c_1, c_2, ..., c_K\}$, where each circuit $c_i$ is composed of a structure $A_i$ and parameters $\theta_i$, the objective is to find the optimal parameters for each circuit as

\begin{equation}
	\theta_i^* = \arg\min_{\theta} \mathcal{L}(c_i, \theta),
\end{equation}

where $\mathcal{L}(c_i, \theta)$ is the loss function for circuit $c_i$. However, as the number of candidate quantum circuits increases, the computational resources required for gradient descent optimization grow exponentially.

To address this issue, parameter reuse techniques have been extensively adopted in QAS. These techniques enhance optimization efficiency by transferring and sharing training parameters among different candidate circuits. The QUEST-A combines parameter reuse with evolutionary algorithms, developing a more adaptive optimization strategy based on this foundation.

Within the evolutionary algorithm framework of QUEST-A, parameter reuse serves as an effective mechanism for knowledge transfer and accumulation across generations. Specifically, in each generation of candidate circuits, the best-performing individual is designated as the focusing agent. After fine-grained trained focusing, its parameters are selectively reused by other individuals in the next generation, thus disseminating valuable knowledge throughout the entire population.

Let $c_{best}$ be the best-performing individual in the current generation as

\begin{equation}
	c_{best} = \arg\max_{c \in \mathcal{P}'_g} E(c),
\end{equation}

where $F(c)$ is the fitness function incorporating both the DAG path count $P(c)$ and circuits expressivity $E(c)$:

\begin{equation}
	Fitness(c) = f(P(c), E(c)).
\end{equation}

The parameter reuse strategy can be formalized as:

\begin{equation}
	\theta_i^{new} = \alpha \theta_{best} + (1-\alpha) \theta_i,
\end{equation}

where $\alpha \in [0,1]$ is the reuse rate, and $\theta_{best}$ are the parameters of $c_{best}$. In actual tasks, $\alpha$ is set to be very close to $1$ to balance the utilization of posteriori knowledge and the diversity of search.

The rationale behind this method is twofold: firstly, based on the characteristics of evolutionary algorithms, individuals within the same generation typically possess similar circuit structures, providing a natural foundation for parameter reuse. Secondly, the re-uploading circuit template adopted by QUEST-A enables almost lossless parameter transfer between structurally similar circuits. This method achieves efficient knowledge transfer between architectures, not only improving search efficiency but also enhancing the algorithm's ability to explore global optima in complex search spaces.

\subsection{Evolutionary Synthesis of Knowledge Transfer}
\label{Evolutionary Synthesis of Knowledge Transfer}

A key process of the QUEST-A framework lies in its integration of the untrained filtering described in Section ~\ref{Untrained Filtering}. and  trained focusing introduced in Section ~\ref{Trained Focusing} into a unified evolutionary algorithm framework. This integration adopts the principles of evolutionary algorithms to guide the QAS process, combining the discrete optimization problem of circuit structures with the continuous optimization of circuit parameters.

In this knowledge-driven search process, untrained filtering and trained focusing work collaboratively, forming an efficient search mechanism. Rapid architecture pruning acts as a preliminary selection mechanism, identifying promising candidate architectures. Meanwhile, focused architecture optimization conducts a refined adaptation process, fully exploiting the potential of each candidate architecture. The alternation of these two processes creates an efficient knowledge transfer cycle, driving the entire search process to continuously evolve towards superior solutions.  The framework of QUEST-A is described in Algorithm ~\ref{Framework}.

\begin{algorithm}
	\caption{QUEST-A Framework}
	\label{Framework}
	\setstretch{1.2}
	\footnotesize
	\begin{algorithmic}[1]
		\STATE Initialize supercircuit $S$ with architecture $A$ and parameters $\theta$
		\STATE Define $Fitness(c) = f(P(c), E(c))$ \COMMENT{Fitness function}
		\FOR{generation $g = 1$ to $G$}
		\STATE $\mathcal{P}_g \leftarrow$ Sample from $S$
		\FOR{$c \in \mathcal{P}_g$}
		\STATE Evaluate $P(c)$ \COMMENT{DAG path count}
		\STATE Evaluate $E(c)$ \COMMENT{circuits expressivity}
		\ENDFOR
		\STATE $\mathcal{P}'_g \leftarrow$ Select top $K$ candidates based on $Fitness(c)$
		\STATE $c_{best} \leftarrow \arg\max_{c \in \mathcal{P}'_g} Fitness(c)$
		\FOR{$c_i \in \mathcal{P}'_g$}
		\STATE $\theta_i^* \leftarrow \arg\min_{\theta} \mathcal{L}(c_i, \theta)$ \COMMENT{Trained focusing}
		\STATE $\theta_i^{new} \leftarrow \alpha \theta_{best} + (1-\alpha) \theta_i^*$ \COMMENT{Parameter reuse}
		\ENDFOR
		\STATE Update $S$ with knowledge from $c_{best}$
		\STATE $\mathcal{P}_{g+1} \leftarrow$ Evolve($\mathcal{P}'_g$) \COMMENT{Crossover and mutation}
		\STATE Update $A$ and $\theta$ in $S$ based on $\mathcal{P}_{g+1}$
		\ENDFOR
		\RETURN $\arg\max_{c \in \mathcal{P}_G} Fitness(c)$
	\end{algorithmic}
\end{algorithm}

In the evolutionary framework of QUEST-A, the quantum circuit structure and parameters collectively constitute the \textit{genes} of individuals, providing a rich space for evolutionary operations. Untrained filtering and trained focusing form the \textit{environment} for individual survival, with DAG path count, KL divergence, and loss function values serving as multi-level fitness evaluation indicators.

This framework achieves a synergy between coarse-grained and fine-grained search processes. Untrained filtering explores broad regions of the search space, while trained focusing refines promising areas, ensuring both search breadth and precision. The population-based evolutionary process allows QUEST-A to maintain and evaluate multiple search regions simultaneously. The supercircuit structure in QUEST-A represents a dynamically evolving search space, where knowledge transfer between generations guides the exploration towards promising architecture regions, enhancing the search efficiency and adaptability. To validate the effectiveness and efficiency of this approach, we now turn to a comprehensive experimental analysis across various quantum computing tasks.

\section{Experiment Results and Performance Analysis}
\label{Experiment Results and Performance Analysis}

\subsection{Experiment Framework}
\label{Experiment Framework}

A series of experiments were designed to validate the effectiveness of QUEST-A's knowledge-driven search strategy. These experiments include: signal representation tasks to demonstrate QUEST-A's performance in generative tasks, particularly in handling complex signals; image classification tasks to verify QUEST-A's performance in practical application scenarios; and variational quantum eigensolver (VQE) tasks to showcase its potential in quantum advantage domains. In each experiment, models were trained five times, with the best performance selected as the comparison result. Detailed information for experiments can be found in ~\ref{Experimental details}.

Numerical simulation experiments were conducted on a Python-based simulation platform, utilizing PennyLane \cite{bergholm2018pennylane} , Mindspore Quantum \cite{xu2024mindspore}, and TensorFlow Quantum \cite{broughton2020tensorflow}. Experiments in a real quantum computing environment were performed on the $\textit{ibm\_kyiv}$ backend.

\subsection{Task \uppercase\expandafter{\romannumeral1}: QUEST-A in signal Representation}
\label{Task I: QUEST-A in signal Representation}

The signal representation task aims to learn the intrinsic representation of signals, with its core objective being the learning of a function that maps input coordinates to corresponding signal values. This task can be viewed as a specialized form of continuous function approximation or signal reconstruction. In the field of quantum computing, this task demonstrates the capability of quantum models in continuous function approximation and also showcases  potential in implicit neural representations tasks \cite{mildenhall2021nerf}.

This experiment utilizes the same audio and image datasets as the benchmark models \cite{zhao2024quantum}, ensuring consistency in data sources and preprocessing methods. Specifically, for the image representation task, images of astronaut, camera, and coffee from the scikit-image dataset \cite{van2014scikit} were selected.These images were first downsampled to a resolution of $32\times 32$ pixels and then resized through a linear layer before being input into a 10-qubit quantum model. The model processed the data through three iterations of data re-uploading for architecture discovery. Additionally, the model is equipped with two variational layers, structurally identical to the re-uploading circuits, both at the initial stage and prior to the measurement stage. These layers serve to optimize the initial state of quantum evolution and the basis vectors for measurement. 

The evolutionary algorithm was configured to run for 10 generations, with each generation comprising 10 individuals. For the dataset $D = {(x_i, y_i)}$, where $x_i$ represents the coordinates of the signal $y_i$, the mean squared error (MSE) is utilized as the performance metric for the model output $y_i\prime$. The fitness function during the evolutionary process is defined as follow

\begin{equation}
	Fitness := -\log\left(\frac{1}{|D|}\sum_{i=1}^{|D|}(y_i-y_i')^2\right).
\end{equation}

\begin{figure}[t]
	\hspace{0cm}
	\includegraphics[width=1\textwidth, trim=0cm 3cm 0.4cm 0.5cm,clip]{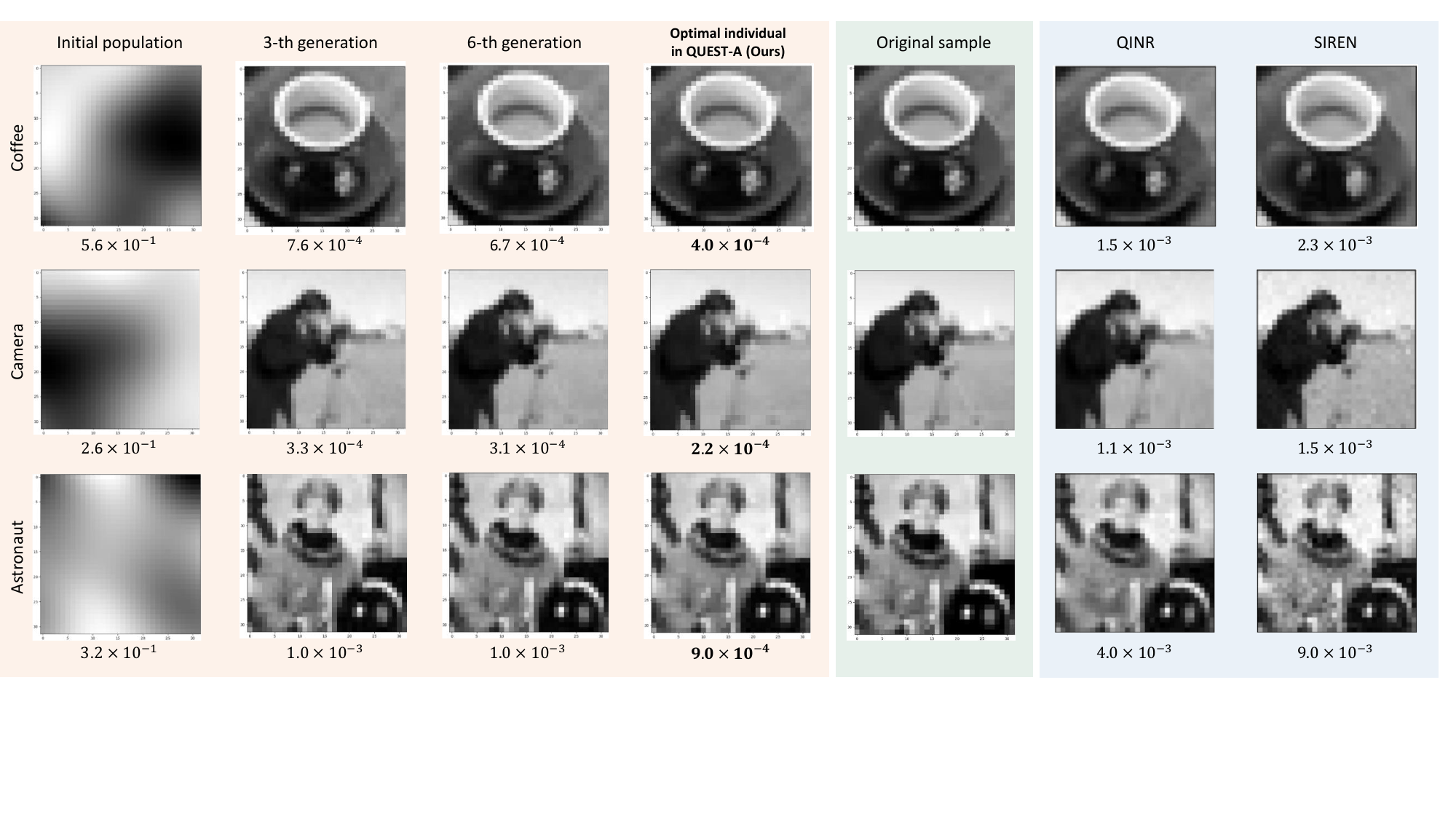}
	\caption{
		\footnotesize
		Performance of QUEST-A in image representation tasks. The numerical values beneath each image represent the corresponding MSE. The red area illustrates the evolution of MSE for the best-performing individual in QUEST-A over generations. The green area shows the original images as the task benchmark. The blue area presents the results and MSE values for baseline \cite{zhao2024quantum,sitzmann2020implicit}.
		}
	\label{image_representation}
\end{figure}

The learning progression and final results of the model on the $\textit{Astronaut}$, $\textit{Camera}, and   $\textit{Coffee} images are visually represented in Fig.~\ref{image_representation}.
 
The results demonstrate that in the initial stages of the QUEST-A method's search process, the candidate circuit population obtained through untrained filtering performed poorly, capturing almost no effective information. However, as knowledge accumulated through iterations, the performance of candidate circuits showed a rapid improvement trend. For the best-performing image, QUEST-A reduced the MSE from $4.0\times10^{-3}$ of the baseline model to $2.2\times10^{-3}$, achieving approximately a $45\%$ reduction in error rate. In the Astronaut image, which exhibits a more complex frequency pattern, QUEST-A similarly achieved a low error rate, demonstrating visually discernible advantages in capturing image details. Compared to the two baseline methods, QUEST-A exhibited superior performance in processing high-frequency information relative to the SOTA quantum methods, resulting in more precise retention of details and edge information in the generated images. The performance of QUEST-A stems from its further enhancement of the inherent advantages of data re-uploading quantum circuits in continuous function approximation.

\begin{figure}[t]
	\hspace{0cm}
	\includegraphics[width=1\textwidth, trim=5cm 10.5cm 0.4cm 0.5cm,clip]{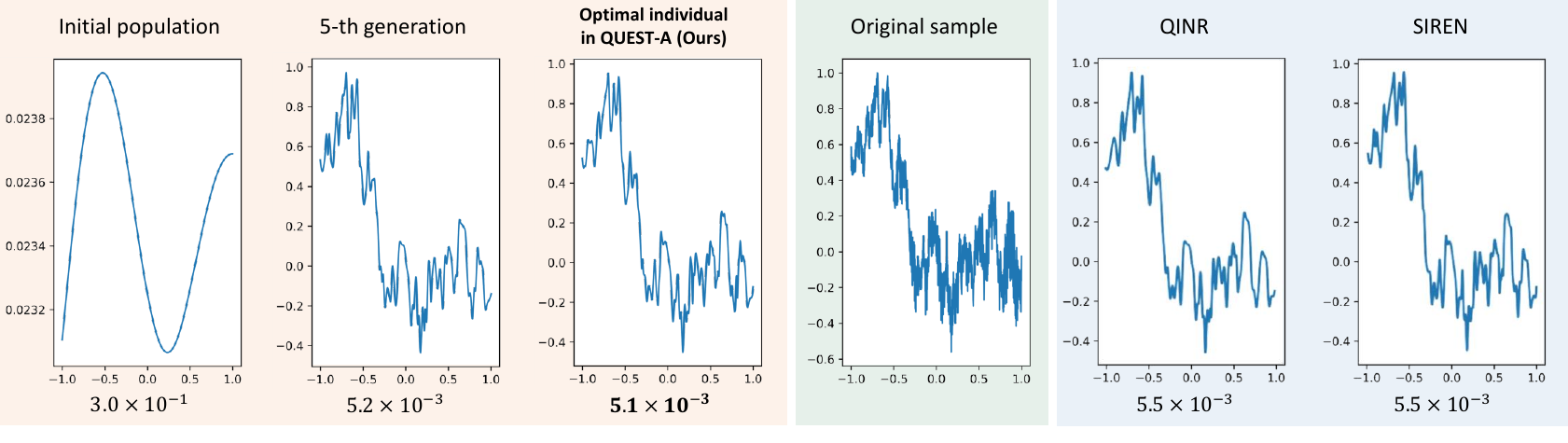}
	\caption{
		\footnotesize
		Performance of QUEST-A in audio representation tasks. The red area illustrates the performance evolution of QUEST-A during the search process and its final results. The green area shows the original audio signals as the task benchmark. The blue area presents the performance of two baseline models on this task.
	}
	\label{audio representation}
\end{figure}

To further validate QUEST-A's capability in processing different types of signals, we also conducted experiments on audio representation tasks, with the results presented in Fig.~\ref{audio representation}.

The results of the audio representation task further confirm the advantages of the QUEST-A model in signal recognition. Although in terms of MSE, the QUEST-A model achieved only a $7.3\%$ reduction in error rate compared to the baseline model, its performance in audio amplitude details processing surpassed state-of-the-art quantum methods. From the perspective of frequency graphs, the results generated by QUEST-A more accurately reproduced the frequency patterns of the benchmark audio, especially demonstrating more refined reproduction in the high-frequency components. While these details may contribute minimally to the overall MSE calculation, they are crucial for the perception of audio quality.

During the training process, we observed that the QUEST-A model quickly learned low-frequency patterns in the initial stages, similar to its performance in the image representation task. In the subsequent training process, the model made fine adjustments to the high-frequency components of the waveform and rapidly converged. This learning pattern further demonstrates QUEST-A's adaptability in handling different types of signal. 

\begin{table}[!t]
	\centering
	\renewcommand{\arraystretch}{0.1}
	\begin{threeparttable}
		\caption{Performance of QUEST-A and SOTA models in signal representation tasks.}
		\scriptsize
		\begin{tabular}{>{\centering\arraybackslash}p{0.12\textwidth} >{\centering\arraybackslash}p{0.11\textwidth} >{\centering\arraybackslash}p{0.11\textwidth} >{\centering\arraybackslash}p{0.115\textwidth} >{\centering\arraybackslash}p{0.115\textwidth} >{\centering\arraybackslash}p{0.115\textwidth} >{\centering\arraybackslash}p{0.115\textwidth}}
			\toprule
			\multirow{2}{*}[-0.5ex]{$\textbf{Method}$} & \multicolumn{2}{c}{$\textbf{Sound Representation}$} & \multicolumn{4}{c}{$\textbf{Image Representation}$} \\[3pt]
			\cmidrule(lr){2-3} \cmidrule(lr){4-7}
			& Cello & params & Astronaut & Camera & Coffee & params\tnote{a} \\[3pt]
			\midrule
			SIREN & 5.5\tnote{b} & 691 & 9.0 & 1.5 & 2.3 & 701 \\[9pt]
			QIREN & 5.5 & 649 & 4.0 & 1.1 & 1.5 & 657 \\[9pt]
			\makecell{$\textbf{QUEST-A}$\\$\textbf{(ours)}$} & \makecell{$\textbf{5.1}$\\$(\textbf{7.3\%$\uparrow$})$} & \makecell{$\textbf{601}$\\$(\textbf{7.4\%$\downarrow$})$} & \makecell{$\textbf{0.9}$\\$(\textbf{77.5\%$\uparrow$})$} & \makecell{$\textbf{0.2}$\\$(\textbf{81.8\%$\uparrow$})$} & \makecell{$\textbf{0.4}$\\$(\textbf{73.3\%$\uparrow$})$} & \makecell{$\textbf{609}$\\$(\textbf{7.3\%$\downarrow$})$}\\
			\bottomrule
		\end{tabular}
		\begin{tablenotes}
			\item[a] The params column indicates the number of parameters used by each model.
			\item[b] Performance is measured by the MSE($\times 10^{-3}$) between the representation results and the task benchmark, lower MSE values indicate better performance.
		\end{tablenotes}
		\label{tab:signal representation}
	\end{threeparttable}
\end{table}

\subsection{Task \uppercase\expandafter{\romannumeral2}: QUEST-A in Image Classification}
\label{Task II: QUEST-A in Image Classification}

Unlike signal representation tasks that focus on pattern discovery, image classification tasks emphasize the search for optimal decision boundaries. This study uses the widely adopted MNIST and Fashion-MNIST datasets as benchmarks, comparing QUEST-A with several state-of-the-art quantum baseline methods, including QUILT \cite{silver2022quilt}, Élivágar \cite{anagolum2024elivagar}, and QNAS \cite{wang2022quantumnas}. To thoroughly evaluate QUEST-A's performance in the NISQ era, we designed three experimental scenarios: numerical simulation, noise simulation, and experiments under real noisy environments. For the simulated noisy environment, we employed a composite noise model incorporating bit flip, phase flip, and depolarizing channel noise. 

In terms of preprocessing, all $28\times28$ pixel images first undergo principal component analysis (PCA), after which the processed data is input into a 10-qubit quantum circuit. The model performs four iterations of data re-uploading and uses cross-entropy as the loss function. The experimental results and performance comparisons on the MNIST and Fashion-MNIST datasets are presented in Tab.~\ref{tab:classification}. 

\begin{table}[t]
	\centering
	\renewcommand{\arraystretch}{0.1}
	\begin{threeparttable}
		\caption{Performance of QUEST-A and SOTA models in classification tasks.}
		\scriptsize
		\begin{tabular}{c ccc cc ccc}
			\toprule
			\multirow{2}{*}[-0.5ex]{\textbf{Task}} & \multicolumn{3}{c}{\textbf{NNS}\tnote{a}} & \multicolumn{2}{c}{\textbf{SNS}} & \multicolumn{3}{c}{\textbf{RNE}} \\
			\cmidrule(lr){2-4} \cmidrule(lr){5-6} \cmidrule(lr){7-9}
			& Quilt & QNAS & \makecell{QUEST-A\\(ours)} & Élivágar\tnote{b} & \makecell{QUEST-A\\(ours)} & QNAS & Élivágar & \makecell{QUEST-A\\(ours)} \\[3pt]
			\midrule
			MNIST-2 & 94.5\tnote{c} & 95.0 & $\textbf{99.9}$ & 97.0 & $\textbf{99.7}$ & 95.0 & 93.0 & $\textbf{98.5}$ \\[10pt]
			MNIST-4 & 80.8 & 77.0 & $\textbf{97.7}$ & 64.0 & $\textbf{90.5}$ & 71.0 & 65.0 & $\textbf{92.0}$ \\[10pt]
			FMNIST-2 & 96.6 & 94.2 & $\textbf{98.1}$ & 97.0 & $\textbf{95.8}$ & 90.0 & 97.0 & $\textbf{94.3}$ \\[10pt]
			FMNIST-4 & 77.7 & 89.0 & $\textbf{91.2}$ & 76.0 & $\textbf{86.0}$ & 85.0 & 73.0 & $\textbf{82.4}$ \\
			\bottomrule
		\end{tabular}
		\begin{tablenotes}
			\item[a] The table presents performance under three conditions: noiseless numerical simulation (NNS), simulated noisy simulation (SNS), and real noisy environment (RNE).
			\item[b] For baseline works, if multiple results were available under different conditions (e.g. different basic gate sets), the best performance was selected as the final performance for the baseline model.
			\item[c] Classification accuracy ($\%$) were obtained through five independent experiments for QUEST-A.
		\end{tablenotes}
		\label{tab:classification}
	\end{threeparttable}
\end{table}

\begin{figure}[h]
	\hspace{0cm}
	\includegraphics[width=0.65\textwidth, trim=0cm 0cm 0cm 0cm,clip]{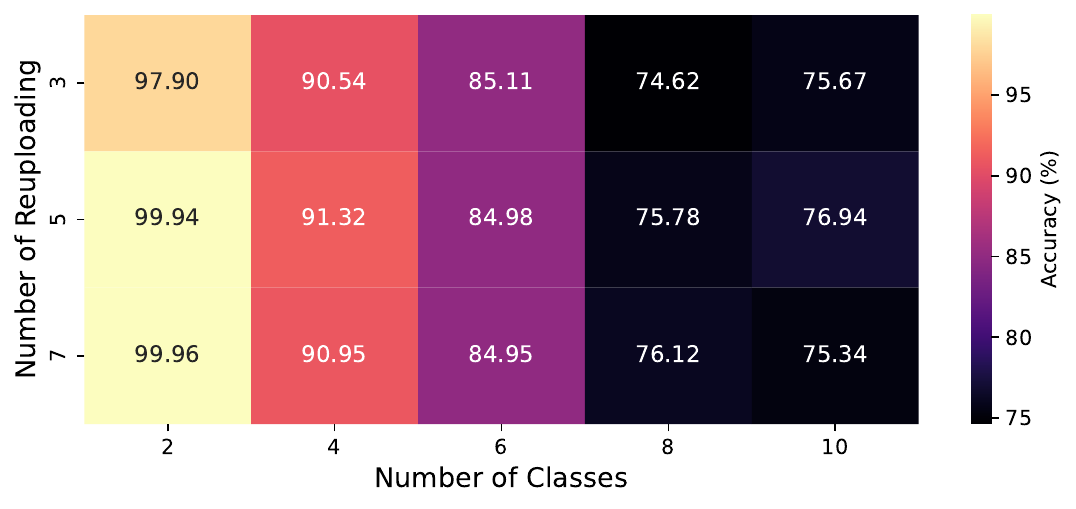}
	\caption{
		\footnotesize
		Heatmap of classification accuracy vs. number of re-uploading. 
	}
	\label{heatmap}
\end{figure}

QUEST-A demonstrated superior search capability and robustness on both MNIST and Fashion-MNIST datasets. Compared to baseline models, QUEST-A achieved higher classification accuracy in noiseless environments (NNS). Notably, when noise was introduced (SNS), QUEST-A's performance showed only a slight decrease. This noise robustness became more evident as task difficulty increased; when transitioning from 2-class to 4-class classification, QUEST-A's performance degradation was less pronounced than that of baseline models. QUEST-A performed well on both MNIST and Fashion-MNIST datasets, indicating its generalization capability and adaptability. Further validation of QUEST-A's performance in classification tasks is presented as Tab.~\ref{tab:multiclass_performance} and Fig.~\ref{heatmap}.

\begin{table}[t]
	\centering
	\renewcommand{\arraystretch}{0.1}
	\begin{threeparttable}
		\caption{Performance of QUEST-A in multi-class classification tasks.}
		\scriptsize
		\begin{tabular}{>{\centering\arraybackslash}p{0.14\textwidth} >{\centering\arraybackslash}p{0.132\textwidth} >{\centering\arraybackslash}p{0.132\textwidth} >{\centering\arraybackslash}p{0.132\textwidth} >{\centering\arraybackslash}p{0.132\textwidth} >{\centering\arraybackslash}p{0.132\textwidth}}
			\toprule
			\multirow{2}{*}[-0.5ex]{\textbf{Dataset}} & \multicolumn{5}{c}{\textbf{Number of Classes}\tnote{a}} \\
			\cmidrule(lr){2-6}
			& 2 & 4 & 6 & 8 & 10 \\[1pt]
			\midrule
			MNIST & \makecell{0.9994\\$(\pm0.3{\times}10^{-3})$} & \makecell{0.9736\\$(\pm3.2{\times}10^{-3})$} & \makecell{0.9391\\$(\pm3.5{\times}10^{-3})$} & \makecell{0.9129\\$(\pm1.7{\times}10^{-3})$} & \makecell{0.8545\\$(\pm2.0{\times}10^{-3})$} \\[14pt]
			FMNIST & \makecell{0.9810\\$(\pm1.3{\times}10^{-3})$} & \makecell{0.9087\\$(\pm4.8{\times}10^{-3})$} & \makecell{0.8513\\$(\pm2.7{\times}10^{-3})$} & \makecell{0.7576\\$(\pm3.5{\times}10^{-3})$} & \makecell{0.7636\\$(\pm2.6{\times}10^{-3})$} \\
			\bottomrule
		\end{tabular}
		\begin{tablenotes}
			\item[a] Performance of QUEST-A on MNIST and Fashion-MNIST for multi-class classification tasks $(\pm$ standard deviation$)$
		\end{tablenotes}
		\label{tab:multiclass_performance}
	\end{threeparttable}
\end{table}

In-depth analysis of image multi-classification tasks further validates the robustness of the QUEST-A model. QUEST-A demonstrated high performance stability across classification tasks ranging from $2$ to $10$ classes. Although model performance showed an expected declining trend as the number of classes increased, QUEST-A still maintained a high accuracy of over $75\%$ in the most challenging 10-class classification task. A noteworthy phenomenon is that increasing the number of trainable layers did not bring about the expected significant performance improvement. On the contrary, in some cases, a slight performance decline was observed. This unexpected result suggests that the model may have experienced a certain degree of overfitting in determining decision boundaries. 

Above experimental results not only further confirm QUEST-A's robustness in handling multi-classification tasks but also provide insights into the model's potential performance in more complex application scenarios.

\subsection{Task \uppercase\expandafter{\romannumeral3}: QUEST-A in Variational Quantum Eigensolver}
\label{Task III: QUEST-A in Variational Quantum Eigensolver}

The VQE task aims to estimate the ground state energy of a given Hamiltonian. In the quantum computing framework, this involves optimizing a parameterized quantum circuit to minimize the expected energy of the system. The VQE task primarily examines the model's ability to prepare complex quantum states, which should closely approximate the ground state of the target Hamiltonian, while also evaluating whether the ansatz design has sufficient expressive power to represent the solution space of the problem.

This experiment selects three typical systems as VQE objects: the $H_2$ molecule (4 qubits), the Heisenberg model (5 qubits), and the Transverse Field Ising Model (TFIM, 6 qubits). Due to the unique characteristics of the VQE, QUEST does not employ data re-uploading circuits in this task. The gate set used for the task includes $RX$, $RY$, $RZ$, and $CNOT$ gates. The Hamiltonian for the $H_2$ molecule follows the construction method from previous work \cite{kandala2017hardware}. The Hamiltonians for TFIM and the Heisenberg model are defined as follows:

\begin{equation}
	H_{TFIM}=\sum_{i=1}^n Z_iZ_{i+1}-X_i,
\end{equation}

\begin{equation}
	H_{Heis}=\sum_{i=1}^nX_iX_{i+1}+Y_iY_{i+1}+Z_iZ_{i+1}+\sum_{i=1}^nz_i,
\end{equation}

where $n$ denotes the number of qubits. The model's performance is measured by the difference between the estimated value $E$ and the theoretically exact ground state energy $E_0$, i.e., $E - E_0$. To ensure fair comparison, QUEST-A and the baseline models use the same number of qubits and similar parameter counts. In each experiment, the models are run multiple times, with the best performance selected as the final result. 

We selected two representative baseline models for comparison, TF-QAS \cite{he2024training} and PQAS \cite{zhang2021neural}, as fast architecture search method. Tab.~\ref{tab:VQE}. presents the performance of QUEST-A and these baseline models on three quantum systems along with detailed comparison results.

\begin{table}[t]
	\centering
	\renewcommand{\arraystretch}{1} 
	\begin{threeparttable}
		\caption{Performance of QUEST-A and SOTA models in VQE tasks.}
		\scriptsize
		\begin{tabular}{>{\centering\arraybackslash}p{0.12\textwidth}  >{\centering\arraybackslash}p{0.12\textwidth}  >{\centering\arraybackslash}p{0.18\textwidth}  >{\centering\arraybackslash}p{0.18\textwidth}  >{\centering\arraybackslash}p{0.20\textwidth}}
			\toprule
			\textbf{Systems} & \textbf{Qubit} & \textbf{PQAS} & \textbf{TF-QAS} & \textbf{QUEST-A (ours)} \\
			\midrule
			$H_2$        & 4     & -\tnote{a}                 & $1.6\times10^{-8}\tnote{b}$ & $\mathbf{5.9\times10^{-15}}$ \\[3pt] 
			Heisenberg   & 5     & $8.9\times10^{-2}$ & $1.2\times10^{-3}$ & $\mathbf{1.7\times10^{-14}}$ \\[3pt] 
			TFIM         & 6     & $9.0\times10^{-2}$ & $1.2\times10^{-5}$ & $\mathbf{2.8\times10^{-8}}$  \\[3pt]
			\bottomrule
		\end{tabular}
		\begin{tablenotes}
			\item[a] The model consistently failed to complete the VQE task for $H_2$ in experimental reproduction.
			\item[b] Model performance is defined by the difference between the model's estimated value and the theoretical exact value.
		\end{tablenotes}
		\label{tab:VQE}
	\end{threeparttable}
\end{table}

QUEST-A achieved the highest precision in all test cases, demonstrating an advantage of up to 11 orders of magnitude over state-of-the-art methods. Compared to similar fast architecture search methods, QUEST-A not only successfully completed the ground state energy prediction task within chemical accuracy $(1.6\times 10^{-3} Ha)$ but also achieved more precise predictions. As the system size increased from $4$ qubits to $6$ qubits, QUEST-A maintained its performance advantage. This outcome reflects QUEST-A's effectiveness in searching complex quantum architectures for ground state approximation. Its exceptional performance may stem from QUEST-A's ability to avoid getting trapped in local optima during the search process, as well as its enhanced sensitivity to the fine structure of quantum states.

Furthermore, the fact that this task did not utilize existing high-performance ansatzes corroborates that QUEST-A's knowledge-driven search strategy is template-independent, but rather can adaptively discover quantum circuit structures suitable for specific problems.

In conclusion, we evaluated QUEST-A's knowledge-driven search strategy across three distinct tasks: insignal representation, image classification, and VQE. In the insignal representation task, QUEST-A achieved up to $81.8\%$ performance improvement compared to SOTA methods, with a simpler architecture space. For image classification, it demonstrated superior search capability and robustness, maintaining over $75\%$ accuracy in $10$-class tasks and showing adaptability to noisy environments. In VQE tasks, QUEST-A outperformed existing methods by up to $11$ orders of magnitude in precision. These results consistently demonstrate the effectiveness of QUEST-A's progressive knowledge learning approach in quantum architecture search.

\section{Discussion}
\label{Discussion}

The QUEST-A framework demonstrates how knowledge-driven search can effectively bridge quantum and classical computing paradigms. While quantum computation excels at rapid estimation and parallel processing, classical optimization techniques provide reliable solutions for parameter tuning and precise optimization. This complementary approach provides guiding principles for developing hybrid algorithms. Another exploration reveals that our retrieval-based approach can discover optimal structures for not only intermediate quantum circuits but also data encoding and measurement schemes. Recent works \cite{wang2023automated,thanos2024automated} in automated circuit design, combined with advances in quantum embedding \cite{havlivcek2019supervised} and measurement \cite{quek2021adaptive}, indicate promising directions for developing task-specific quantum architectures.

Furthermore, the flexibility of the QUEST-A framework allows for incorporation of hardware-specific factors. However, caution must be exercised when considering hardware noise in QAS. Although many current works \cite{wang2022quantumnas,anagolum2024elivagar,lu2023qas,he2024meta} consider quantum hardware characteristics such as qubit architecture and noise properties, the time-varying nature of quantum noise challenges the effectiveness of static search strategies. This suggests the need for more adaptive knowledge accumulation mechanisms in noisy environments, which could be a key focus for future QAS.

\section{Conclusion}
\label{Conclusion}

In this work, we propose QUEST-A, a knowledge-driven framework for QAS through alternating phases of untrained filtering and trained focusing. This approach views QAS as an item retrieval task with dynamic knowledge accumulation, decomposing the search process into coarse-grained retrieval guided by circuit inherent properties and fine-grained retrieval guided by task performance. Experiments demonstrate QUEST-A's performance across signal representation, image classification, and VQE tasks, validating the effectiveness of progressive knowledge learning in architecture searching.

Our exploration reveals three key insights into quantum architecture search. First, QAS is essentially a knowledge accumulation process that starts with limited prior understanding and discovers optimal architectures through continuously enriching the knowledge of circuit characteristics. Second, validated circuit templates serve as crucial prior knowledge that determines the initial scope and efficiency of the search process. Third, effective QAS requires dynamic knowledge-driven strategies rather than static search methods, where the process progressively learns and transfers architectural knowledge, analogous to knowledge distillation in modern deep learning.

Future work should focus on developing knowledge-driven search strategies in QAS, particularly exploring the design of quantum-principle-guided circuit templates as efficient search starting points.

\section*{Authors' contributions}
\textbf{Lian-Hui Yu}: Conceptualization, Data curation, Methodology. \textbf{Xiao-Yu Li}: Methodology, Project administration. \textbf{Geng Chen}: Formal analysis, Investigation, Writing – original draft. \textbf{Qin-Sheng Zhu}: Formal analysis, Validation. \textbf{Hui Li}: Methodology, Software. \textbf{Guo-Wu Yang}: Software.

\section*{Availability of supporting data}
All data generated or analysed during this study are available and included in this published article. \href{https://github.com/UESTC-YLH/QUEST-A}{Code} for our work has been open-sourced.

\section*{Acknowledgements}
This work was supported by the Open Fund of Advanced Cryptography and System Security Key Laboratory of Sichuan Province (Grant No. SKLACSS-202210), National Natural Science Foundation of China (Grant No. 62472072 and No. 62172075).

\small
\bibliographystyle{elsarticle-harv}
\bibliography{quslref}

\newpage

\appendix 

\section{Preliminaries of Quantum Circuit}
\label{appendixA}

The essence of quantum computing lies in embedding data into Hilbert space and processing information using quantum properties such as superposition and entanglement. Within this framework, the role of quantum basic gates is to ensure that the operable domain fully covers the target Hilbert space, enabling precise control and manipulation of quantum states.

In QAS task, basic gates are the fundamental units of circuit design. The core challenge lies in discovering optimal architectures through progressively understanding the characteristics and combinations of these basic gates. The basic gates used in this paper include $R_X$, $R_Y$, $R_Z$, and $CNOT$ gates, whose mathematical forms are as follows: 

\begin{equation}
RX(\theta) = \begin{pmatrix} \cos(\theta/2) & -i\sin(\theta/2) \\ -i\sin(\theta/2) & \cos(\theta/2) \end{pmatrix}, 
RY(\theta) = \begin{pmatrix} \cos(\theta/2) & -\sin(\theta/2) \\ \sin(\theta/2) & \cos(\theta/2) \end{pmatrix},
\end{equation}

\begin{equation}
	RZ(\theta) = \begin{pmatrix} e^{-i\theta/2} & 0 \\ 0 & e^{i\theta/2} \end{pmatrix}, CNOT = \begin{pmatrix} 1 & 0 & 0 & 0 \\ 0 & 1 & 0 & 0 \\ 0 & 0 & 0 & 1 \\ 0 & 0 & 1 & 0 \end{pmatrix}.
\end{equation}

where, CNOT is a two-qubit gate, while the others act on single qubits. Similar to NAS, the challenge of QAS stems from the exponentially growing search space due to gate combinations and task-specific search objectives.

\section{Experimental details}
\label{Experimental details}

The experimental settings for our work are presented in the Tab.~\ref{tab: Detailed parameter}

\begin{table}[t]
	\centering
	\renewcommand{\arraystretch}{1} 
	\begin{threeparttable}
		\caption{Detailed parameter settings for all experiments in our work}
		\scriptsize
		\begin{tabular}{c ccc}
			\toprule
			\textbf{Parameter} & \textbf{\makecell{Signal\\Representation}} & \textbf{\makecell{Image\\Classification}} & \textbf{VQE} \\
			\midrule
			Number of qubits      & 8                      & 10                   &\makecell{ 5 ($Heisenberg$)\\4 ($H_2$), 6 ($TFIM$)}       \\[8pt]
			Data re-uploading     & 2                      & 3                    & -\tnote{a}                                 \\[7pt]
			Number of generations  & 10                     & 10                   & 10                                \\[7pt]
			Population size        & 5                      & 5                    & 5                                 \\[7pt]
			Loss metric            & MSE                    & Crossentropy         & -                                 \\[7pt]
			\makecell{Trainable parameters\tnote{b}}   & \makecell{601 ($Sound$)\\ 609 ($Image$)} & 175                  & \makecell{175 ($Heisenberg$)\\105 ($H_2$), 420 ($TFIM$)} \\[7pt]
			\bottomrule
		\end{tabular}
		\begin{tablenotes}
			\item[a] The VQE task does not utilize specific circuit templates, nor does it employ gradient-based optimization processes.
			\item[b] Experimental parameters are set to be comparable with baseline models.
		\end{tablenotes}
		\label{tab: Detailed parameter}
	\end{threeparttable}
\end{table}

In the search process, the best-performing individual is selected as the final performance metric for the experiment. All experiments were conducted independently five times, with the best performance across these runs reported as the final result. The \href{https://github.com/UESTC-YLH/QUEST-A}{code} for QUEST-A is available.

\section{Ablation Analysis}
\label{Ablation Analysis}

We conducted a set of ablation experiments to validate the effectiveness of progressive knowledge learning using the Fashion-MNIST 10-class classification task as the experimental scenario. The experiments investigated the following three cases: ({\romannumeral1}) Trained focusing in random circuits . ({\romannumeral2})  Trained focusing after DAG-based untrained filtering only; ({\romannumeral3}) Trained focusing after DAG and KL-based untrained filtering;

For random trained focusing (case {\romannumeral1}), we used $5$ randomly generated circuits for focusing. For the untrained filtering cases (cases {\romannumeral2} and {\romannumeral3}), we screened from $100$ candidate circuits and selected the top $5$ optimal circuits.  All experiments were terminated upon convergence of the loss function.

\begin{figure}[H]
	\hspace{0cm}
	\includegraphics[width=0.85\textwidth, trim=0cm 0cm 0cm 0cm,clip]{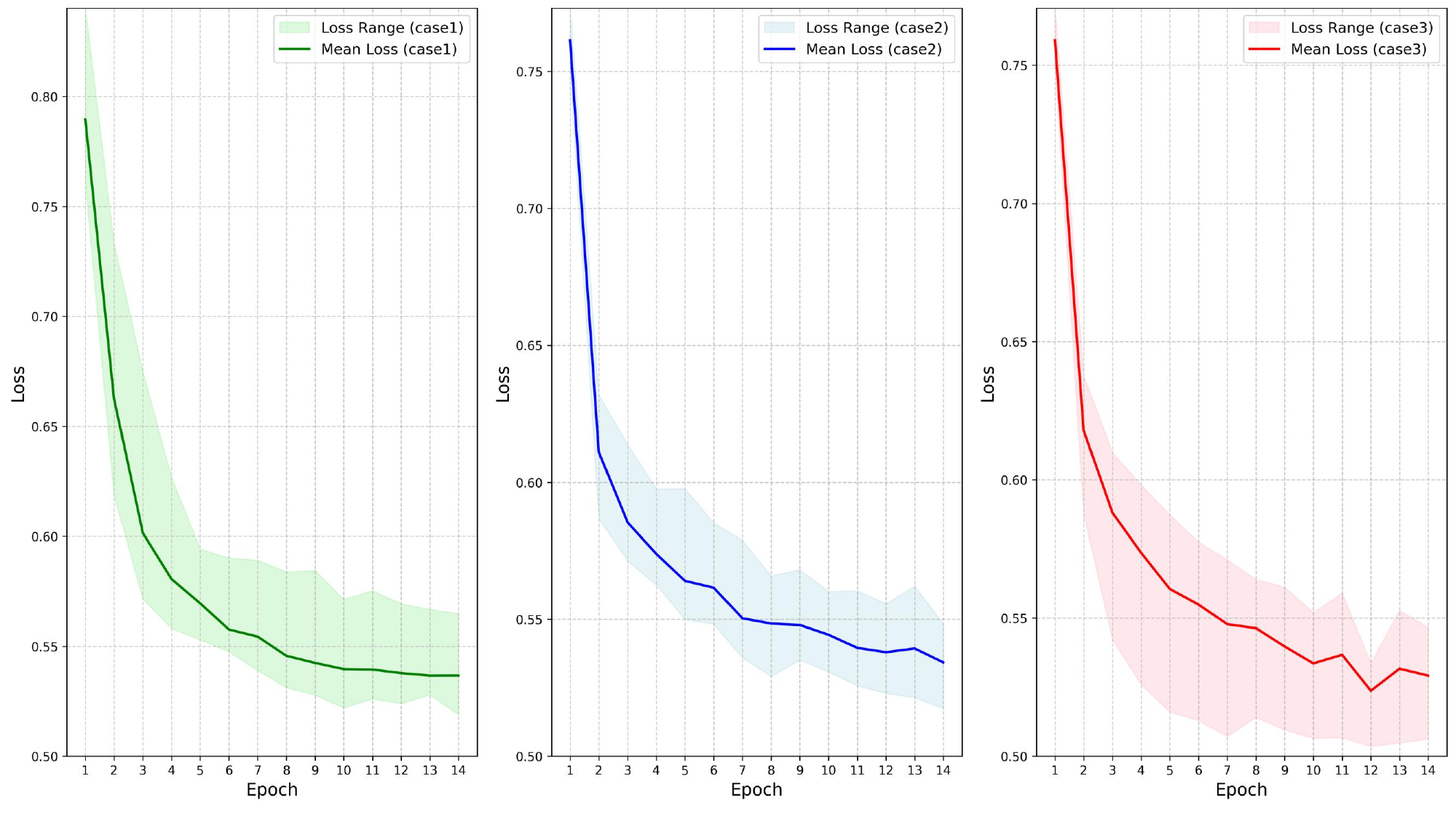}
	\caption{
		\footnotesize
		Loss trend over epochs for three case. Training progress shown until convergence of case {\romannumeral1}.
	}
	\label{Ablation_Analysis}
\end{figure}

Case ({\romannumeral3}) with complete untrained filtering exhibited faster convergence and superior performance compared to the other two cases. Case ({\romannumeral1}) with random circuits demonstrated slower convergence and inferior final performance. Comparing case ({\romannumeral1}) with cases using untrained filtering, it is evident that the filtering process effectively enhanced the initial search space and convergence capability, indicating better architecture discovery potential. Moreover, the comprehensive untrained filtering process in case ({\romannumeral3}) versus case ({\romannumeral2}) yielded further improvements in both convergence speed and final performance.

\end{document}